\documentclass[12pt,a4paper]{article}

\usepackage{amsmath}
\usepackage{amsfonts}
\usepackage{amssymb}
\usepackage{graphicx}
\usepackage{color}
\usepackage{booktabs}
\usepackage{inputenc}
\usepackage[T1]{fontenc}
\usepackage{mathrsfs}
\usepackage{enumerate}
\usepackage{xcolor,url}
\usepackage{cancel,dsfont} 

\pdfoutput=1

\newcommand{\eea}{\end{eqnarray}}
\newcommand{\bea}{\begin{eqnarray}}

\def\be{\begin{equation}}
\def\ee{\end{equation}}

\usepackage[colorlinks,bookmarks]{hyperref}
\definecolor{linkblue}{rgb}{0,0,0.8}
\definecolor{linkgreen}{rgb}{0,0.5,0}

\hypersetup{pdfpagemode=UseNone, pdfstartview=FitH, linkcolor=linkblue,
            citecolor=linkgreen, urlcolor=linkblue}

\begin{document}

\begin{center}

{\Large \bf {The Hubble Tension
in Light of \\[0.3cm]
 the Full-Shape Analysis of  Large-Scale Structure Data}
}
\\[0.7cm]

{\large  Guido D'Amico${}^{1}$,  Leonardo Senatore${}^{2,3}$,  Pierre Zhang${}^{4,5,6}$, Henry Zheng${}^{2,3}$\\[0.7cm]}

\end{center}

\begin{center}

\vspace{.0cm}

{\normalsize { \sl $^{1}$ Dipartimento di SMFI dell' Universita' di Parma \& INFN Gruppo Collegato di Parma, Parma, Italy}}\\
\vspace{.3cm}

{\normalsize { \sl $^{2}$ Stanford Institute for Theoretical Physics, Physics Department,\\ Stanford University, Stanford, CA 94306}}\\
\vspace{.3cm}

{\normalsize { \sl $^{3}$ 
Kavli Institute for Particle Astrophysics and Cosmology,\\
 SLAC and Stanford University, Menlo Park, CA 94025}}\\
\vspace{.3cm}

{\normalsize { \sl $^{4}$ Department of Astronomy, School of Physical Sciences, \\
University of Science and Technology of China, Hefei, Anhui 230026, China}}\\
\vspace{.3cm}

{\normalsize { \sl $^{5}$ CAS Key Laboratory for Research in Galaxies and Cosmology, \\
University of Science and Technology of China, Hefei, Anhui 230026, China}}\\
\vspace{.3cm}

{\normalsize { \sl $^{6}$ School of Astronomy and Space Science, \\
University of Science and Technology of China, Hefei, Anhui 230026, China}}\\
\vspace{.3cm}

\vspace{.3cm}

\end{center}

\hrule \vspace{0.3cm}
{\small  \noindent \textbf{Abstract} The disagreement between direct late-time measurements of the Hubble constant from the SH0ES collaboration, and early-universe measurements based on the $\Lambda$CDM model from the Planck collaboration might, at least in principle, be explained by new physics in the early universe. Recently, the application of the Effective Field Theory of Large-Scale Structure to the full shape of the power spectrum of the SDSS/BOSS data has revealed a new, rather powerful,  way to measure the Hubble constant and the other cosmological parameters from Large-Scale Structure surveys. In light of this, we analyze two models for early universe physics, Early Dark Energy and Rock 'n' Roll, that were designed to significantly ameliorate the Hubble tension. Upon including the information from the full shape to the Planck, BAO, and Supernovae measurements, we find that the degeneracies in the cosmological parameters that were introduced by these models are well broken by the data, so that these two models do not significantly ameliorate the tension. 

\noindent

\vspace{0.3cm}}
\hrule

\vspace{0.3cm}
\newpage

\tableofcontents

\section{Introduction and Summary\label{sec:intro}}

\paragraph{The EFTofLSS:} After a long development, the Effective Field Theory of Large-Scale Structure (EFtofLSS) has been applied to the data of the BOSS/SDSS survey, and used to analyze the  power spectrum of galaxies~\cite{DAmico:2019fhj,Ivanov:2019pdj,Colas:2019ret}~\footnote{{Notice that Ref.~\cite{Colas:2019ret} is a companion paper to~\cite{DAmico:2019fhj}.} Ref.~\cite{DAmico:2019fhj} also applied it to the bispectrum, but finding marginal improvements, probably due to the fact that only the tree-level prediction was being used, so that the $k$-reach was not quite high.}.
These results have allowed us to measure, using only a prior from Big Bang Nucleosynthesis~(BBN), all the cosmological parameters of the $\nu\Lambda$CDM model and, recently, also of the $w$CDM model~\cite{DAmico:2020kxu}, except for the neutrino masses and for $w$, for which only bounds have been obtained.
The $w$ parameter is well measured upon adding Baryon Acoustic Oscillation (BAO) data.
The smallness of the error bars on some of these parameters, {and the accuracy achieved when fitting to simulations,} have shown the power of current Large-Scale Structure (LSS) surveys even without the inclusion of any priors from the cosmic microwave background~(CMB).
In particular, the precision on the measurement of the present-day dark matter fraction, $\Omega_m$, is very similar to the one from Planck2018~\cite{Aghanim:2018eyx}; and the precision on the present-day Hubble parameter, $H_0$, is of the same order of the one obtained from the Cosmic Distance Ladder, such as SH0ES~\cite{Riess:2019cxk}; the constraint on $w$ is not far from the one obtained by CMB plus supernovae~\cite{DAmico:2020kxu}. 
These results establish that the contribution of next generation LSS surveys to our  understanding of the history of the universe, once analyzed with a controlled theory such as the EFTofLSS, might be very large, potentially helping humankind to continue the remarkable cosmological exploration that was achieved in the past decades.

It was a very long journey to develop the EFTofLSS to such a level that it could be applied to data: each of the ingredients of the EFTofLSS that was required in order to be able to analyze the data  (dark matter and baryon clustering, galaxy clustering, IR-resummation, redshift space distortions, etc.)  was singly developed, tested on simulations, and shown to be successful.
Though not all these intermediate results are directly used in the analysis, we believe they were {\it necessary} for us, and probably for anybody else, to apply the model to data. We therefore find it fair, in each instance where the EFTofLSS is applied to data, to add the following footnote where we acknowledge at least a fraction of those crucially important developments~\footnote{The initial formulation of the EFTofLSS was performed in Eulerian space in~\cite{Baumann:2010tm,Carrasco:2012cv}, and subsequently extended to Lagrangian space in~\cite{Porto:2013qua}. The dark matter power spectrum has been computed at one-, two- and three-loop orders in~\cite{Carrasco:2012cv, Carrasco:2013sva, Carrasco:2013mua, Carroll:2013oxa, Senatore:2014via, Baldauf:2015zga, Foreman:2015lca, Baldauf:2015aha, Cataneo:2016suz, Lewandowski:2017kes,Konstandin:2019bay}. These calculations were accompanied by some  theoretical developments of the EFTofLSS, such as a careful understanding of renormalization~\cite{Carrasco:2012cv,Pajer:2013jj,Abolhasani:2015mra} (including rather-subtle aspects such as lattice-running~\cite{Carrasco:2012cv} and a better understanding of the velocity field~\cite{Carrasco:2013sva,Mercolli:2013bsa}), of several ways for extracting the value of the counterterms from simulations~\cite{Carrasco:2012cv,McQuinn:2015tva}, and of the non-locality in time of the EFTofLSS~\cite{Carrasco:2013sva, Carroll:2013oxa,Senatore:2014eva}. These theoretical explorations also include an enlightening study in 1+1 dimensions~\cite{McQuinn:2015tva}. An IR-resummation of the long displacement fields had to be performed in order to reproduce the Baryon Acoustic Oscillation (BAO) peak, giving rise to the so-called IR-Resummed EFTofLSS~\cite{Senatore:2014vja,Baldauf:2015xfa,Senatore:2017pbn,Lewandowski:2018ywf,Blas:2016sfa}.  An account of baryonic effects was presented in~\cite{Lewandowski:2014rca}. The dark-matter bispectrum has been computed at one-loop in~\cite{Angulo:2014tfa, Baldauf:2014qfa}, the one-loop trispectrum in~\cite{Bertolini:2016bmt}, and the displacement field in~\cite{Baldauf:2015tla}.
The lensing power spectrum has been computed at two loops in~\cite{Foreman:2015uva}.  Biased tracers, such as halos and galaxies, have been studied in the context of the EFTofLSS in~\cite{ Senatore:2014eva, Mirbabayi:2014zca, Angulo:2015eqa, Fujita:2016dne, Perko:2016puo, Nadler:2017qto} (see also~\cite{McDonald:2009dh}), the halo and matter power spectra and bispectra (including all cross correlations) in~\cite{Senatore:2014eva, Angulo:2015eqa}. Redshift space distortions have been developed in~\cite{Senatore:2014vja, Lewandowski:2015ziq,Perko:2016puo}. Neutrinos have been included in the EFTofLSS in~\cite{Senatore:2017hyk,deBelsunce:2018xtd}, clustering dark energy in~\cite{Lewandowski:2016yce,Lewandowski:2017kes,Cusin:2017wjg,Bose:2018orj}, and primordial non-Gaussianities in~\cite{Angulo:2015eqa, Assassi:2015jqa, Assassi:2015fma, Bertolini:2015fya, Lewandowski:2015ziq, Bertolini:2016hxg}. Faster evaluation schemes for the calculation of some of the loop integrals have been developed in~\cite{Simonovic:2017mhp}. Comparison with high-quality $N$-body simulations to show that the EFTofLSS can accurately recover the cosmological parameters have been performed in~\cite{DAmico:2019fhj,Colas:2019ret,Nishimichi:2020tvu}}.

\paragraph{The Hubble Tension:} There are roughly three different ways in which the present value of the Hubble constant, $H_0$, is measured (see~\cite{Verde:2019ivm} for a recent review).
The first that was historically developed is the direct measurement based on the cosmic distance ladder.
Depending on the elements of the ladder that are chosen (the so-called calibration method), there are currently two measurements available, either from the SH0ES collaboration~\cite{Riess} or from the Carnegie-Chicago Hubble Program (CCHP)~\cite{Freedman:2019jwv}.
A second way is to use the measurement of the density fluctuations to extract the angle upon which the scale associated to the horizon at the last scattering surface is projected, and from there to extract the Hubble parameter.
This has been historically applied first to the CMB data~(see for e.g.~\cite{Aghanim:2018eyx}).
Recently, these measurements have been shown to be possible also with Large-Scale Structure data within a small range in redshift~\cite{DAmico:2019fhj,Ivanov:2019pdj, Colas:2019ret,DAmico:2020kxu,Philcox:2020vvt}, giving competitive results~\footnote{For earlier results that involve using BAO measurements at more spaced redshift, see~\cite{Addison:2017fdm}. In~\cite{Abbott:2017smn} the galaxy lensing is also included.}.
Finally, a third way to measure $H_0$ is to use the time delay in multiply-imaged gravitational lenses, as shown for example by the H0LiCOW collaboration~\cite{Wong:2019kwg}. 

Roughly, and within the $\Lambda$CDM model, the measurements involving CMB and LSS show a value of $H_0$ which is significantly lower than the ones obtained from cosmic ladder and gravitational lenses.
If we consider the most extreme, but also the most precise, result from the SH0ES collaboration, the tension is larger than four standard deviations~\footnote{The tension with the measurement from the CCHP collaboration~\cite{Freedman:2019jwv} is significantly smaller, if not absent at all.}.
This is the so-called `Hubble tension'.

There are two crucial differences between the CMB and LSS measurements and the ones from the cosmic ladder and gravitational lenses. 
First, the CMB and LSS measurements obtain their value of $H_0$ by assuming some physical law to be valid in the early universe: from the Big Bang all the way to present data {(though only the time from around the last scattering surface onwards really matters)}.
Because of this, these are called ``early universe'' measurements.
On the contrary, the cosmic ladder and the gravitational lenses need to assume physical laws of the universe only near the present time, and one could argue that they are much less sensitive to these assumptions.
Because of this, these are called ``late universe'' measurements or also ``direct'' measurements.

The second aspect in which the early universe and the late universe methods differ is the complexity of the physical system one needs to model in order to extract the measurement. While the measurements from both the CMB and the LSS (using the EFTofLSS) involve the solution of linear or mildly non-linear equations, the late universe measurements require the modeling of complex astrophysical systems which are described by very non-linear equations and that, in most cases, {are modeled phenomenologically rather than solved from first principles}. 

Because of this second difference between the two classes of measurements, it is well possible that the disagreement in the Hubble measurements might be due to some systematic error either in the measurements or in the astrophysical modeling in either of the methods.
It is not the purpose of this paper to investigate this possibility~\footnote{{The disagreement between the two main collaborations using SNIa, the SH0ES~\cite{Riess:2019cxk} and the CCHP~\cite{Freedman:2019jwv}, may point to some unresolved systematics in this kind of measurements. A criticism of a possible oversimplified modeling of lenses has been presented in~\cite{Kochanek:2019ruu}.}}, though we make two observations. First, the CMB and LSS measurements are sensitive to different systematic and modeling errors, and, therefore, the agreement between these two measurements found in~\cite{Aghanim:2018eyx,DAmico:2019fhj,Ivanov:2019pdj,Colas:2019ret} suggests that these issues are small in the two datasets. Second, the measurement from cosmic ladder from the CCHP collaboration~\cite{Freedman:2019jwv} is compatible, albeit maybe marginally, with the measurements from CMB and LSS. 

There is a second possibility for the explanation of the disagreement in the $H_0$ determination. Because of the first difference between the ``early'' and ``late'' Universe measurements, it is actually possible that a model of the early universe beyond $\Lambda$CDM might change the early universe dynamics so that the inferred parameter values from the CMB and LSS can shift to agree with the late-universe measurements.
Several models have been proposed to realize this possibility, and it is fair to say that, when analyzed using data sets that included SH0ES data, the tension was ameliorated (but not solved)~\footnote{We mention here that a critique has been presented in~\cite{Krishnan:2020obg} on the class of models that change early universe dynamics, based on results from fits to $H_0$ using several low-redshift distance indicators.}.
Two particularly popular models are the so-called Early Dark Energy model (EDE)~\cite{Poulin:2018cxd,Smith:2019ihp}  and the Rock 'n' Roll (RnR) model~\cite{Agrawal:2019lmo}, that use a scalar field to inject energy in the early universe, raising the inferred value of $H_0$ from CMB and LSS measurements.

The purpose of this paper is to perform the analysis of several cosmological data sets against the EDE and RnR models, and to see the  amount to which, if any, the Hubble tension is still alleviated upon inclusion in the analysis of the Full Shape (FS) of the BOSS power spectrum. Recently, Ref.~\cite{Hill:2020osr} has analyzed the EDE model against several data sets, but not the BOSS FS.
They have found that  upon adding to the combination of the Planck and SH0ES data other LSS data sets, such as for example BAO~\cite{Alam:2016hwk} or the DES~\cite{Abbott:2017wau} and KiDS~\cite{Hildebrandt:2016iqg} data, or Supernovae, the tension in the Hubble parameter is still persistent and actually increasing even in the EDE model. 
Furthermore, the same reference has found that using just the Planck data is sufficient to determine the EDE parameters, and the resulting values are inconsistent with the one obtained upon adding SH0ES: this suggests that the SH0ES data and the Planck data are inconsistent even in the EDE model.
For the case of the EDE model, we will agree with the findings of~\cite{Hill:2020osr}, and we will make the conclusions stronger by adding the FS data.

\paragraph{Data Sets:} In this paper we focus on applying the EFTofLSS to the Full Shape (FS) of the Power Spectrum of Galaxies to constrain the RnR and EDE models.
We analyze various combinations of data, among which the FS of BOSS DR12 pre-reconstructed power spectrum measurements \cite{Gil-Marin:2015sqa}, baryon acoustic oscillations (BAO) of BOSS DR12 post-reconstructed power spectrum measurements \cite{Gil-Marin:2015nqa}, Planck2018 TT,TE,EE+lowE + lensing~\cite{Aghanim:2018eyx}. We also consider combinations with Supernovae (SN) measurements from the Pantheon Sample \cite{Scolnic:2017caz}, and, finally, the direct measurement of the Hubble constant from the SH0ES collaboration~\cite{Riess}.
When quoting BAO, we also include measurements at small redshift from 6DF~\cite{Beutler:2011hx} and SDSS DR7 MGS \cite{Ross:2014qpa}, as well as high-redshift Lyman-$\alpha$ forest auto-correlation and cross-correlation with quasars from eBOSS DR14 measurements \cite{Agathe:2019vsu, Blomqvist:2019rah}~\footnote{The inclusion of post-reconstructed BAO measurements gives a non-negligible improvement because the reconstruction amounts to using higher $n$-point functions. However the pre- and post-reconstruction BAO measurements are correlated. This is taken into account as in~\cite{DAmico:2020kxu} {(see also~\cite{Philcox:2020vvt})}.}. When combined with Planck or SN, we simply add the {log-}likelihoods, since all the measurements refer to separate redshift bins. There is a small cross-correlation between the galaxy clustering data and the Planck weak lensing, which we neglect.

\paragraph{Main Results:} The main results of our analysis are best represented by Fig.~\ref{fig:rnr_summary} and \ref{fig:ede_summary}. We find that both for the RnR and EDE models, the Planck+BAO+SN data are sufficient to determine the parameters of the models and these are quite inconsistent with the ones obtained when adding the SH0ES data. Upon adding the FS data, we find that, without SH0ES data, the cosmological parameters become more compatible with the ones  obtained from $\Lambda$CDM (implying no evidence for any of the models), while, upon adding SH0ES data, the value of $H_0$ is still low. All these statements are made significantly stronger by the addition of the FS data. The improvement of the goodness of the fit to the data in the EDE and RnR models with respect to $\Lambda$CDM is marginal for all experiments but for SH0ES, where the improvement is significantly decreased upon adding the FS.  In summary, even in the case of EDE and RnR models, the SH0ES data seems to be incompatible with the Planck+FS+BAO+SN data, and, upon inclusion of the SH0ES data, the $H_0$ tension is not resolved, but just very mildly ameliorated by these models.

We also provide a physical explanation for why the EDE and RnR models are not able to resolve the Hubble tension. This has to do with the fact that the data, even without the SH0ES measurement, are able to break the degeneracy between the sound horizon at recombination, $r_s$, and $H_0$. So, ultimately $r_s$ and $H_0$ are not allowed to move much away from their $\Lambda$CDM values, even if the model would allow for this to happen. It is expected therefore that similar findings as here will be found for different models that attempt to use the degeneracy between $r_s$ and $H_0$ to resolve the $H_0$ tension. 

\paragraph{Future directions:} It would be interesting to repeat a similar analysis on different models that claim to ameliorate the Hubble tension, as for example~\cite{Park:2019ibn,Kreisch:2019yzn}.
It would also be interesting to repeat the analysis by removing at least a fraction of the data sets, to see if some model significantly reduces the tension within the resulting subset of data, as it was recently done for Planck, by removing the high multipole data, in~\cite{Chudaykin:2020acu}.

\paragraph{Codes:}  The predictions for the FS of the galaxy power spectrum in the EFTofLSS are obtained using PyBird: Python code for Biased tracers in ReDshift space, publicly available at \href{https://github.com/pierrexyz/pybird}{https://github.com/pierrexyz/pybird} and described in~\cite{DAmico:2020kxu}.
The linear power spectrum in the RnR model has been computed using the ${\rm CLASS}_{-}$RNR code, made available to us by the authors of~\cite{Agrawal:2019lmo} and now publicly available at \href{https://github.com/mwt5345/class_ede}{https://github.com/franyancr/class$_{-}$rnr}, while in the EDE model it has been computed using the code CLASS$_{-}$EDE publicly available at \href{https://github.com/franyancr/class$_{-}$rnr}{https://github.com/mwt5345/class$_{-}$ede} and described in~\cite{Hill:2020osr} (we also compared with the results obtained using AxiCLASS, publicly available at \href{https://github.com/PoulinV/AxiCLASS}{https://github.com/PoulinV/AxiCLASS}, finding agreement). The linear power spectra {in the $\Lambda$CDM model} were computed with the CLASS Boltzmann code~\cite{Blas_2011}~\footnote{ \href{http://class-code.net}{http://class-code.net}}.
The structure of the PyBird code is such that it is practically immediate to interface it with any Boltzmann code.
The posteriors were sampled using MontePython cosmological parameter inference code~\cite{Brinckmann:2018cvx, Audren:2012wb}~\footnote{ \href{https://github.com/brinckmann/montepython\_public}{https://github.com/brinckmann/montepython\_public}}.
The plots have been obtained using the GetDist package~\cite{Lewis:2019xzd}.

\section{Mini-review of Models}

\subsection{Physical Considerations}
\label{sec:review}

Given the current tension between the local determination of the $H_0$ constant and the global fits to the $\Lambda$CDM model, there have been many theoretical models proposed in order to alleviate the tension.
A good review of the possible directions is~\cite{Knox:2019rjx}, where it is argued that the class of models ``least unlikely to be successful'' involves an increase in the expansion $H(z)$ prior to recombination, in order to decrease the sound horizon at recombination.
This comes from the fact that the CMB constrains with precision the angular acoustic scale at recombination $\theta_s$, which is the ratio
\begin{align}\label{eq:thetas}
  \theta_s = \frac{r_s(z_{\rm CMB})}{D_A(z_{\rm CMB})}
\end{align}
where $z_{\rm CMB} \approx 1100$ is the recombination redshift, $r_s = \int^{\infty}_{z_{\rm CMB}} \frac{dz'}{H(z')}c_s(z')$ is the sound horizon at recombination, and $D_A(z) = \int^{z}_{0}\frac{dz'}{H(z')}$ the angular diameter distance.
$D_A$ is sensitive to physics after the CMB decoupling and therefore has explicit dependence on $H_0$, while $r_s$ depends on physics before recombination.
To keep $\theta_s$ constant, an increase in Hubble of $\mathcal{O}(10\%)$ before $z_{\rm CMB}$ allows for a $\mathcal{O}(10\%)$ increase in $H_0$, sufficient to resolve the $H_0$ tension.
A way to achieve this is to add light degrees of freedom that behave like dark energy prior to recombination, and becoming a subdominant component afterwards, to provide consistency with the CMB spectrum~\cite{axiverse, Poulin:2018cxd, Agrawal:2019lmo, Smith:2019ihp, Kaloper:2019lpl, Alexander:2019rsc, Lin:2019qug, Niedermann:2019olb,Sakstein:2019fmf, Niedermann:2020dwg}.
Here we focus on two models, Axion Early Dark Energy (EDE) and Rock'n'Roll Dark Energy (RnR).

Before moving on to review the models and actually perform the data analysis, it is worthwhile to elaborate on the dependence of eq.~(\ref{eq:thetas}) on the cosmological parameters, which explains why adding the LSS data will be powerful in constraining the EDE and RnR models.
Keeping in mind that the relative amplitude of the CMB and BAO peaks fixes $\omega_b$, following for example~\cite{Percival:2002gq,Poulin:2018cxd} we can Taylor expand eq.~(\ref{eq:thetas}) around the best fit given by the Planck cosmology, to find that the dependence of this angle on the remaining cosmological parameters is
\be\label{eq:thetacmb}
\theta_s(z_{\rm CMB})\sim r_s \omega_{m}^{0.4} H_0^{0.2} \ ,
\ee

In LSS we are measuring the same angle $\theta_s(z)$, but with $z\approx z_{\rm LSS}\sim 0.3$. Interestingly, in the case of LSS, the actual angle (or ratio of length scales) $\theta_{\rm LSS}$, under which the BAO oscillation is observed, scales as the geometric mean of what is observed parallel and perpendicular to the line of sight (see for example~\cite{Eisenstein:2005su,Percival:2007yw}):
\be\label{eq:BAOtotal}
\theta_{\rm LSS}\simeq \frac{ r_s(z_{\rm CMB})}{\left(D_A(z_{\rm LSS})^2\cdot c\, z_{\rm LSS}/H(z_{\rm LSS})\right)^{1/3}} \, ,
\ee
where  $c$ is the speed of light. Taylor expanding around the Planck cosmology, we find that $\theta_{\rm LSS}$ scales as~\cite{DAmico:2019fhj}:
\be
\theta_{\rm LSS}\sim r_s \omega_{m}^{0.1} H_0^{0.8} \ . 
\ee

Next, the absolute amplitude of the peaks can be estimated following~\cite{Mukhanov:2003xr} to scale as
\be\label{eq:peaks}
{\rm Peak\; height}\sim r_s^{-0.26} \omega_m^{-0.25}\ .
\ee  
It is straightforward to check at this point that all the parameters $r_s$, $H_0$ and $\omega_m$ can be independently measured by the combination of these three observations, as we will show explicitly in App.~\ref{app:shift}. Of course, there are also other sources of information from CMB and LSS, such as the fact that we observe $\theta(z_{\rm LSS})$ at several redshifts, or the shape of the smooth part of the dark matter power spectrum~\cite{Mukhanov:2003xr,DAmico:2019fhj}. The smooth part of the power spectrum starts playing a role, together with the peak height, only once the FS is analyzed, and it is particularly important to determine $\omega_m$ through its dependence on the equality time. We discuss this more in App.~\ref{sec:lcdm_comp}.

But suffice it to say that, at least in principle, given a certain shift in $r_s$, the shift in $H_0$ needed in the CMB to match observations will be different than the one needed to match the LSS observations, introducing a tension where currently there is no tension. This is, in essence, the reason why the FS and the BAO data will be useful to constrain the EDE and RnR models. More in general, since these considerations apply to the fact that the model-induced degeneracy between $r_s$ and $H_0$ is actually broken by the data, it is expected that similar findings will apply to other models that attempt to use this potential degeneracy to resolve the Hubble tension.

\subsection{(Axion) Early Dark Energy}

The axion EDE model~\cite{Poulin:2018cxd, Smith:2019ihp} introduces an ultra-light axion (ULA) field $\phi$ with the Lagrangian~\cite{axiverse},
\begin{align}
  \mathcal{L} = -\frac{1}{2}(\partial_{\mu}\phi)^2 - \Lambda^4\left(1-\cos \left(\phi/f\right)\right)^n - V_{\Lambda}\ ,
\end{align}
where $f$ is the axion field decay constant, and the misalignment angle $\Theta = \phi/f$ takes values $\Theta \in \left[ -\pi, \pi \right]$.
Denoting $V_n(\phi) = \Lambda^4(1-\cos \phi/f)^n + V_{\Lambda}$, the equation of motion is
\begin{align}
  \ddot{\phi} + 3H\dot{\phi} + \frac{dV_n(\phi)}{d\phi} = 0 \, ,
\end{align}
and the energy density $\rho_{\phi}$ and pressure $p_{\phi}$ are
\begin{equation}
  \rho_{\phi} = \frac{\dot{\phi}^2}{2} + V_n(\phi) \, , \qquad
  p_{\phi} = \frac{\dot{\phi}^2}{2} - V_n(\phi) \, .
\end{equation}
At early times, $H \gg m$, where $m = \Lambda^2 / f$, the axion field is frozen and the energy density grows with respect to the other components.
As Hubble drops below $m$, the field becomes dynamical and begins to approach the minimum of the potential.
The dynamics is parameterized by $m$, $f$, $n$, $V_{\Lambda}$, and $\Theta_i$ where $\Theta_i$ is the
initial misalignment angle.
Without loss of generality, we can restrict $\Theta_i$ to $0 \leq \Theta_i \leq \pi$.

The axion field gives maximum energy injection around the redshift $z_c$ at which it begins to oscillate, that is $m \approx 3 H(z_c)$.
The maximum energy injection, $f_{\textrm{EDE}}$, is given by
\begin{align}
  f_{\textrm{EDE}}(z_c) \simeq \frac{V_n(\Theta_i)}{3M^2_pH^2}\mid_{z = z_c} \simeq \frac{m^2f^2}{3M^2_pH^2(z_c)}(1-\cos \Theta_i)^n \simeq \frac{3 f^2}{M^2_p}(1-\cos \Theta_i)^n \, .
\end{align}
Therefore we can trade the free parameters of the model with the phenomenological parameters $f_{\textrm{EDE}}$, $z_c$, $n$, $\Theta_i$ and $V_\Lambda$ \cite{Smith:2019ihp, Hill:2020osr}.

After $z_c$, the field oscillates and the energy density rapidly decays.
The oscillations occur about the local minimum of the potential $V_n(\phi)$, which to first order is $V \sim \phi^{2n}$.
Using the virial theorem, the equation of state becomes \cite{turner}
\begin{align}
  w_n \equiv \frac{n-1}{n+1} \, .
\end{align}
One should note, however, that if the potential becomes too steep, that is if $n$ is too large, then the field may reach an attractor solution and never oscillates \cite{peebles,liddle}.

\subsection{Rock 'n' Roll}
To increase $H(z)$ in a small redshift window before recombination, one can seek solutions such that the difference of the equation of state of the scalar field $w$ and the one of the background $w_b$, $(w - w_b)$, must transition from negative to positive.
At early times, the scalar field is frozen and the energy density is dominated by the background, so $(w - w_b)$ is negative.
At the redshift where $w = w_b$, the energy density of the field reaches its maximum.
As $w$ increases further, the energy density of the field then becomes subdominant again.
In the Rock 'n' Roll~(RnR) model~\cite{Agrawal:2019lmo} one considers a simple class of solutions with $w \approx -1$
initially, while then $w$ thaws to a constant $w_{\phi} > w_b$.
The redshift $z_c$ at the transition point $w = w_b$ is the point of maximum energy injection.

Afterwards, the energy density of a rolling field with constant equation of state is
\begin{align}
  \rho_{\phi}(a) = \rho_0\left( \frac{a_0}{a} \right)^{3(1 + w_{\phi})}\ ,
\end{align}
where $\rho_0$ is the initial energy density of the rolling solution at a scale factor $a_0$.
Using the Friedmann equations, one finds that
\begin{align}
  V(\phi) &= \frac{1-w_{\phi}}{2}\rho_{\phi} \, , \\
  \partial_a\phi &= \frac{\sqrt{(1 + w_{\phi}\rho_{\phi})}}{aH} \, .
\end{align}
Since the background energy density dominates, $3H^2M^2_P \approx \rho_b = \rho_{b0}\left( \frac{a_0}{a}
\right)^{3(1 + w_b)}$, the solution for $\phi(a)$ is then
\begin{equation}
  \phi(a) = c\left( \frac{a_0}{a} \right)^{3(w_{\phi} - w_b)/2}, \, \qquad
  c = \frac{M_P}{(w_{\phi} - w_b)}\sqrt{\frac{4(1 +
  w_{\phi})\rho_0}{3\rho_{b0}}} \, ,
\end{equation}
imposing that $\phi \to 0$ for $a \to \infty$.
Since both the potential and the field are power laws of $a$, the potential is a power-law of~$\phi$:
\begin{equation}
  V(\phi) = \frac{1}{2}(1 - w_{\phi})\rho_0\left( \frac{\phi}{c} \right)^{2n}, \, \qquad
  n = \frac{1 + w_{\phi}}{w_{\phi} - w_b} \, .
\end{equation}

The potential for the RnR model is therefore chosen as
\begin{equation}
  V(\phi) = V_0\left( \frac{\phi}{M_P} \right)^{2n} + V_{\Lambda} \, ,
\end{equation}
where a constant $V_{\Lambda}$ is added.
As mentioned, the field achieves its maximum energy density when $w = w_b$ at $z_c$, which is given by $\rho_{\phi}(z_c) \approx V(\phi_i)$, where $\phi_i$ is the initial field value.
The maximum energy fraction is given by \cite{Agrawal:2019lmo},
\begin{equation}
  f_{\textrm{RNR}}(z_c) \approx \frac{V(\phi_i)}{3H^2(z_c)M^2_P} \, .
\end{equation}
The free parameters of the RnR model are $V_0$, $V_{\Lambda}$, $n$, and the initial condition $\phi_i$.
{We set $\dot{\phi}_i = 0$ because of the frozen field attractor solution at early times.}
As we see, we can trade two of them for the phenomenological parameters $z_c$, $f_{\textrm{RNR}}$.

\section{Data Analysis}

\subsection{RnR Analysis}
\label{sec:RnR}
\begin{figure}[ht]
  \centerline{\includegraphics[scale = 0.60]{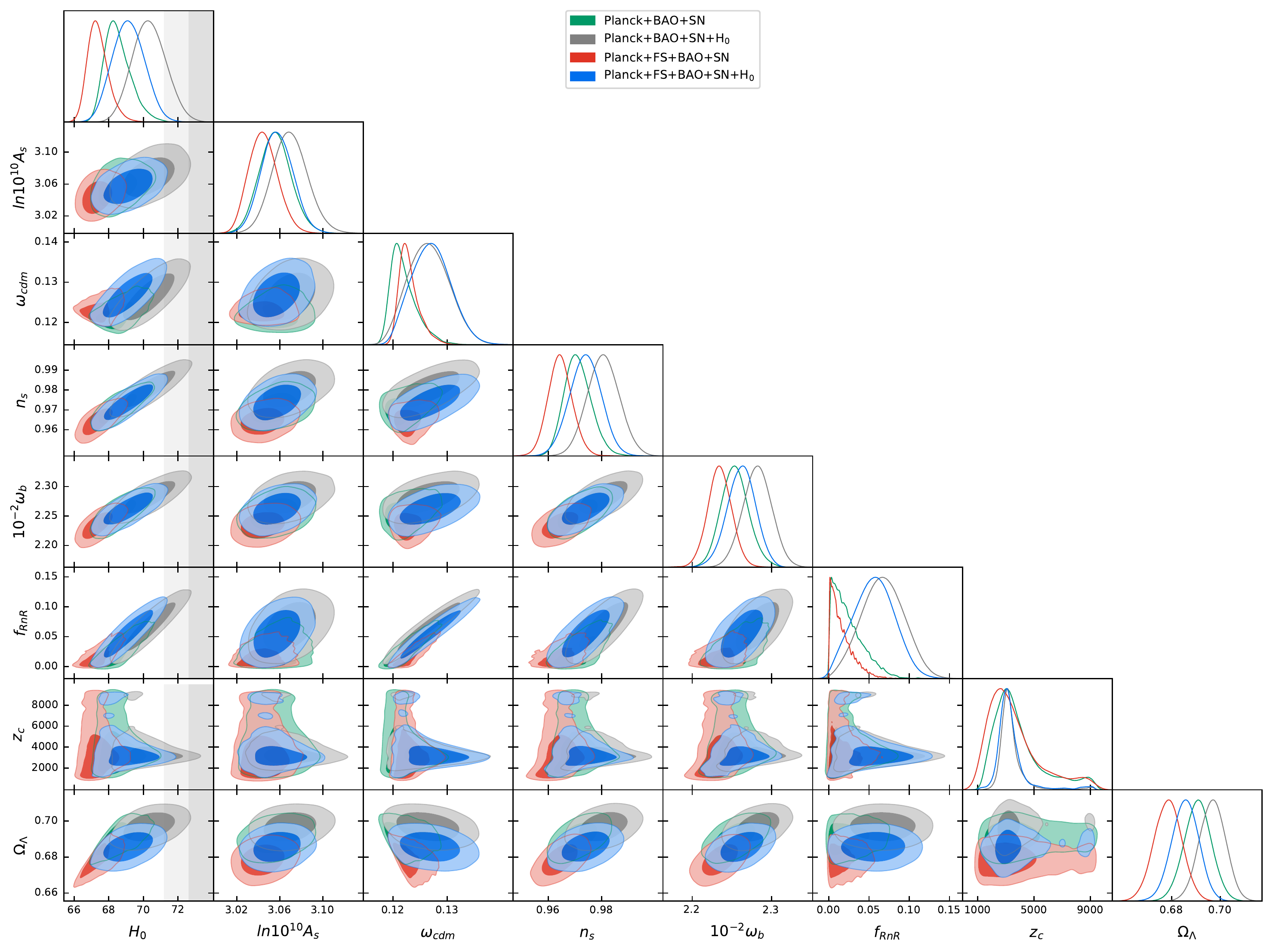}}
  \caption{One dimensional and two dimensional posterior distributions of some of the parameters of the RnR model analyzed with several combinations of data sets: Planck+BAO+Pantheon, Planck+FS+BAO+Pantheon, Planck+BAO+Pantheon+SH0ES and Planck+FS+BAO+SN+SH0ES.}
  \label{fig:rnr_summary}
\end{figure}

For the cosmological analysis of the RnR model, we fix the initial $\dot\phi$ to vanish, and take as free parameters $\phi_i$, $V_0$, and $V_\Lambda$. By
reparameterization, we vary $z_c$, $f_{\rm RnR}$ and $V_\Lambda$.
We further restrict the analysis to the case of $n = 2$ as~\cite{Agrawal:2019lmo} finds that $n=2$ provides the largest values of energy injection and best overall fit.

We show the likelihood contours in Fig.~\ref{fig:rnr_summary}, and in Table~\ref{tab:RNR} we summarize the one-dimensional posteriors for some parameters.
In Tables~\ref{tab:PlanckRnR} and~\ref{tab:PlanckFSRnR} we give the $\chi^2$ of the best fit of our results, compared with the $\Lambda$CDM model run with the same experiments.
We discuss potential issues related to volume sampling of the distribution in App.~\ref{app:sampling}, finding that they give marginal effects.

The RnR model clearly points to a tension between the SH0ES data and all the others.
Both sets of experiments that include SH0ES are best fitted by an injection of scalar field energy, quantified by the fraction $f_{\rm RnR} \sim 0.05$ around {redshift $3400$}.
The $H_0$ best fit is higher than in the corresponding sets of experiments without the SH0ES measurement included.

The models without the SH0ES measurement included have $f_{\rm RnR}$ peaked towards $0$, indicating no dark energy injection in the history of the universe, which confirms that the CMB and LSS datasets are consistent, and that the parameters of the RnR model are determined by them in such a way that the scalar field contribution is not needed. This points to an incompatibility between the Planck and BOSS data versus the SH0ES data, because the values of $H_0$ that one gets before combining the experiments are incompatible.

An interesting point is that adding the FS measurements shifts $H_0$ towards the Planck $\Lambda$CDM value, pointing once again to the consistency of the LSS and Planck data, and to the tension with the SH0ES determination of $H_0$. This happens both with and without the addition of the SH0ES data.
More precisely, without the SH0ES measurement, adding the FS to Planck+BAO+SN results in a shift on $H_0$ by about $1.2 \sigma$, in the direction of the concordance $\Lambda$CDM model.
For the measurements with SH0ES, adding the FS to Planck+BAO+SN results in a shift on $H_0$ by about $0.9 \sigma$, in the direction of the concordance $\Lambda$CDM model~\footnote{In evaluating the significance of these shifts, we remind that in these analyses a large fraction of the data sets, as well as the way they are analyzed, are unchanged.}.
In App.~\ref{sec:lcdm_comp} we show the posteriors for the  parameters that are common with the $\Lambda$CDM model for the combinations of the data sets not involving SH0ES, showing indeed consistency of the posteriors.

Moreover, in the RnR model there is not a clear consistency with the SH0ES result either, even when including the SH0ES data.
Taking the results without the FS, we have a residual tension with SH0ES of about $2.15 \sigma$ when comparing the $H_0$ best fit~\footnote{The number of $\sigma$'s for the tension with respect to the SH0ES (or Planck) measurement is computed the following way.
We approximate the posteriors from each experiment as a Gaussian, and then consider the distribution of the difference of the $H_0$ parameter, which is a Gaussian whose mean is the difference of the means and whose variance is the sum in quadrature of the variances. The effective number of $\sigma$'s is defined as the ratio of the mean and the standard deviation of the resulting distribution.}.
Adding the FS, the residual tension increases to $2.9 \sigma$, an improvement from the $4.2 \sigma$ without SH0ES, but hardly a resolution of the tension.

We can finally give a qualitative explanation for why the RnR is not ultimately able to resolve the $H_0$ tension, given the freedom in changing $r_s$ from the $\Lambda$CDM value. As anticipated by the physical considerations in sec.~\ref{sec:review}, in App.~\ref{app:shift} in Fig.~\ref{fig:rs_plot} we show that in this model, even without the inclusion of the SH0ES measurements, $r_s$ is well measured by the combination of Planck+FS+BAO+SN data to be close to the $\Lambda$CDM value, leaving not much freedom to adjust $H_0$ towards the SH0ES value.
In the same figure, one can also see the increase in the precision from the inclusion of the FS information.

\begin{table}
\begin{center}
\begin{tabular}{lllll}
  \toprule
  Dataset & $H_0$ & $f_{\rm scf}$ & $\sigma_{\rm SH0ES}$ &  $\sigma_{\rm Planck}$ \\
  \midrule
  Planck + BAO + SN &  $68.52^{+0.55}_{-0.89}$ & $<0.062\ (95\% {\rm \; CL})$ & 3.5 & 1.3\\
  \midrule
  Planck + FS + BAO + SN& $67.39^{+0.46}_{-0.68}$ & $<0.045\ (95\% {\rm \; CL})$ & 4.3 & 0.0 \\
  \midrule
  Planck + BAO + SN + SH0ES & $70.34 \pm 0.97$ & $0.067 \pm 0.026$ & 2.1 & 2.7 \\
  \midrule
  Planck + FS + BAO + SN + SH0ES & $69.14 \pm 0.92$ & $0.056 \pm 0.025$ & 2.9 & 1.6\\
  \bottomrule
\end{tabular}
\end{center}
\caption{Mean and $1$-$\sigma$ intervals for the $H_0$ and $f_{\rm scf}$ parameters in the RnR model,  as well as the effective number of $\sigma$'s that the Hubble measurement is away from either SH0ES and Planck.}
\label{tab:RNR}
\end{table}

\subsection*{Conclusions for the RnR Analysis}
\label{sec:RnR_conclusion}
We summarize the following conclusions from our analysis above:
\begin{itemize}
  \item The model-independent direct measurement of $H_0$ from the late-universe from the SH0ES collaboration gives $H_0 = 74.03\pm1.42\;{\rm km\,s}^{-1} {\rm Mpc}^{-1}$~\cite{Riess}.
  From Table~\ref{tab:RNR}, without the constraint on $H_0$ from SH0ES, the Planck+FS+BAO+SN datasets determine an $H_0$ that is in tension with the SH0ES measurement.
  If the FS dataset is not included, the tension with the SH0ES determination of $H_0$ is $3.5\sigma$, while with the FS inclusion the tension increases to $4.3\sigma$. 

  \item Datasets without the SH0ES constraint have the  scalar field energy injection $f_{\rm RnR}$ approaching zero.
  $f_{\rm RnR}$ tends more strongly towards zero when FS is included.
  {This points to the fact that the $\Lambda$CDM model is a robust fit to these data.}

  \item The \textit{Planck} collaboration determined $H_0 = 67.36 \pm 0.54\, {\rm km}\, {\rm s}^{-1} {\rm Mpc}^{-1}$ within the $\Lambda$CDM model~\cite{Aghanim:2018eyx}. The inclusion of the FS shifts $H_0$ towards the Planck $\Lambda$CDM determination of $H_0$.
  Without the SH0ES constraint on $H_0$, the shift is $1.2\sigma$ and with the SH0ES constraint the shift is $0.9\sigma$.

  \item Even after the constraint on $H_0$ from SH0ES is added, the $H_0$ posterior is still in tension with the SH0ES measurement of $H_0$.
  Without the FS, the tension is $2.15\sigma$ and with the FS the tension increases to $2.9\sigma$.

  \item As we show in Tables~\ref{tab:PlanckRnR} and~\ref{tab:PlanckFSRnR}, the $\chi^2$ of the fit does not improve with respect to $\Lambda$CDM for any experiment but for SH0ES.
  When we add the FS data, the improvement is much decreased.
\end{itemize}
 
\subsection{EDE Analysis}
\label{sec:EDE}
\begin{figure}[!h]
  \centerline{\includegraphics[scale = 0.55]{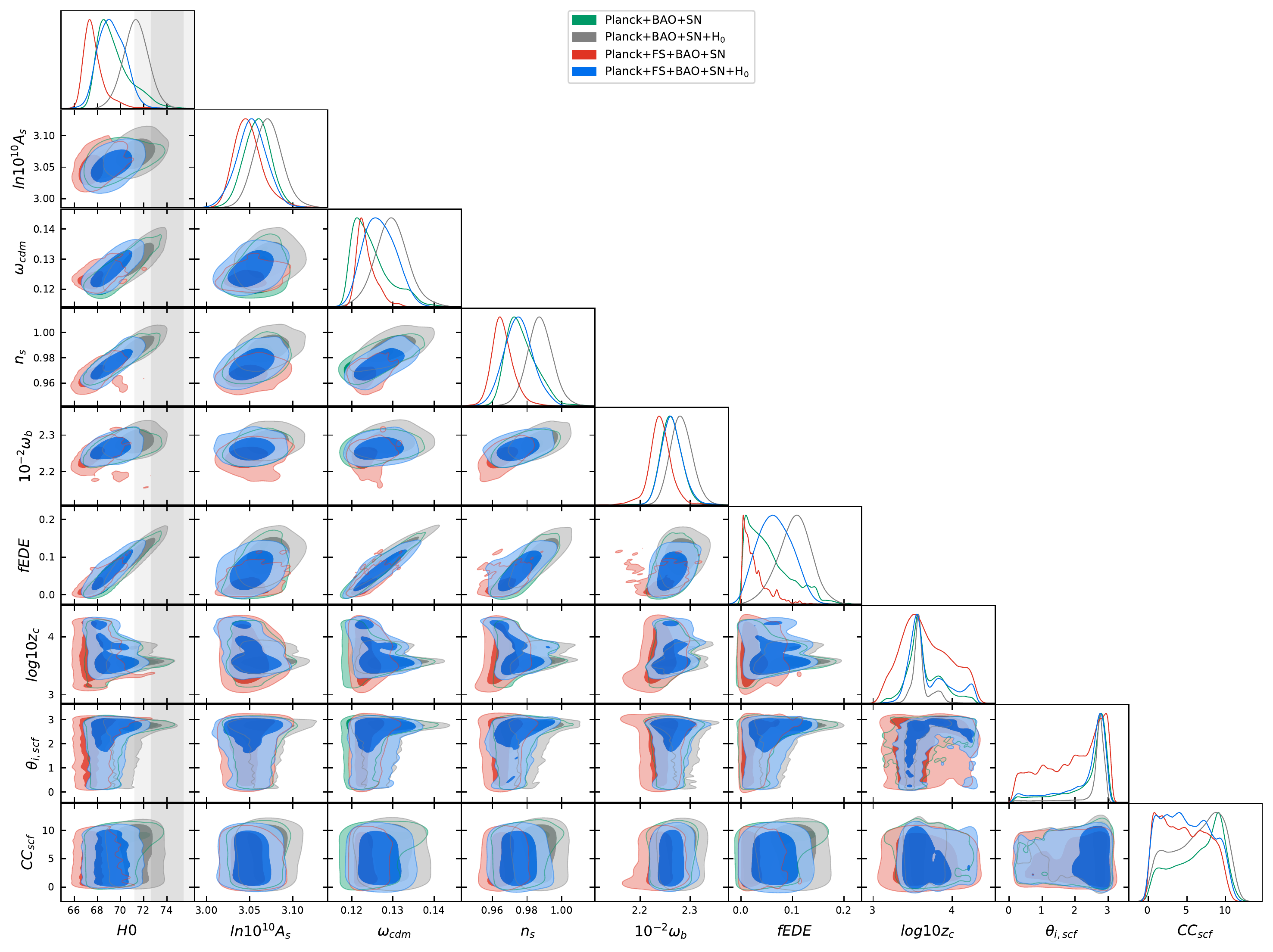}}
  \caption{One dimensional and two dimensional posterior distributions of some of the parameters of the EDE model analyzed with several combinations of data sets: Planck+BAO+Pantheon, Planck+FS+BAO+Pantheon, Planck+BAO+Pantheon+SH0ES and Planck+FS+BAO+SN+SH0ES. Here we define $\theta_{i, scf} = \theta_i$ and $CC_{scf} = V_{\Lambda}$.}\label{fig:ede_summary}
\end{figure}

We now move on to analyze the EDE model, where we will find similar results as for the RnR model.
We restrict our analysis to the case $n = 3$, the best fit integer value reported in \cite{Smith:2019ihp}.
The free parameters of the model are $\theta_i$, $\log(z_c)$, $f_{\rm EDE}$ and $V_{\Lambda}$. 

We show the likelihood contours in Fig.~\ref{fig:ede_summary}, and in Table~\ref{tab:EDE} we summarize the one-dimensional posteriors for some parameters.
In Tables~\ref{tab:PlanckRnR} and~\ref{tab:PlanckFSRnR} we give the $\chi^2$ of the best fit of our results, compared with the $\Lambda$CDM model run with the same experiments.
We discuss potential issues related to volume sampling of the distribution in App.~\ref{app:sampling}, finding that they give marginal effects.
Moreover, we discuss potential issues due to consistency of the BOSS and Planck datasets in App.~\ref{app:Alss}, again finding that our results are robust.

Although the chain convergence is a bit worse than in the RnR model {(the EDE has one more free parameter)}, the results are similar.
Both sets of experiments that include SH0ES are best fitted by an injection of scalar field energy, quantified by the energy fraction $f_{\rm EDE} \sim 0.1$ around redshift $z_c\sim 4000$.
The $H_0$ best fit is higher than in the corresponding sets of experiments without the SH0ES measurement included.

The analyses without the SH0ES measurement have $f_{\rm EDE}$ peaked towards $0$, indicating no dark energy injection in the history of the universe, which confirms that the CMB and LSS datasets are consistent, and that the parameters of the EDE model are determined by them in such a way that the scalar field contribution is not needed.

An interesting point is that adding the FS data shifts $H_0$ towards the Planck $\Lambda$CDM value, pointing once again to the consistency of LSS and Planck data, and to a tension with the SH0ES determination of $H_0$.
In App.~\ref{sec:lcdm_comp} we indeed show the posteriors for the cosmological parameters that are common with the $\Lambda$CDM model, showing consistency of the posteriors.

When including the SH0ES data, the tension with SH0ES (in the case without the FS) is decreased to about $1.5 \sigma$, a notable improvement with respect to the vanilla $\Lambda$CDM model of $4.2 \sigma$.
Adding the FS, the residual tension increases to $2.6 \sigma$, still an improvement from the $\Lambda$CDM model of $4.2 \sigma$, but this points more the consistency of LSS and Planck data than a resolution of the tension with SH0ES.

Finally, the physical reason for why the EDE model is not able to resolve the tension is the same as for the RnR model. As anticipated by the theoretical considerations in sec.~\ref{sec:review}, in App.~\ref{app:shift} we show that the value of $r_s$ is well determined to be close to the $\Lambda$CDM value already by the Planck+FS+BAO+SN data, so that the model cannot exploit this model-induced degeneracy between $r_s$ and $H_0$ to move the inferred value of $H_0$.

\begin{table}
  \begin{center}
  \begin{tabular}{lllll}
    \toprule
    Dataset & $H_0$ & $f_{\rm EDE}$ & $\sigma_{\rm SH0ES}$ &  $\sigma_{\rm Planck}$ \\
    \midrule
    Planck + BAO + SN &  $69.45^{+0.72}_{-1.8}$ & $<0.14\ (95\% {\rm \; CL})$ & 2.4 & 1.5 \\
    \midrule
    Planck + FS + BAO + SN& $68.57^{+0.48}_{-1.0}$ & $<0.08\ (95\% {\rm \; CL})$ & 3.4 & 1.3 \\
    \midrule
    Planck + BAO + SN + SH0ES & $71.3 \pm 1.2$ & $0.104^{+0.034}_{-0.029}$ & 1.5 & 3.0\\
    \midrule
    Planck + FS + BAO + SN + SH0ES & $69.2^{+1.1}_{-1.2}$ & $0.066^{+0.033}_{-0.036}$ & 2.6 & 1.4\\
    \bottomrule
  \end{tabular}
  \end{center}
  \caption{Mean and $1$-$\sigma$ intervals for the $H_0$ and $f_{\rm EDE}$ parameters in the EDE model, as well as the effective number of $\sigma$'s that the Hubble measurement is away from either SH0ES and Planck.}
  \label{tab:EDE}
\end{table}

\subsection*{Conclusions for the EDE Analysis}
\label{sec:EDE_conclusions}
From our analysis, we draw the following conclusions about the EDE model:
\begin{itemize}
  \item From Table~\ref{tab:EDE}, without the SH0ES constraint on $H_0$, both Planck+BAO+SN and Planck+FS+BAO+SN datasets obtain values of
  $H_0$ that are in tension with the SH0ES measurement. The tension with the SH0ES determination of $H_0$ for the analysis without FS information is about $2.4\sigma$, which grows to $4.0 \sigma$ when the FS dataset is included.
  
  \item Without the SH0ES constraint on $H_0$, $f_{\rm EDE}$ tends towards zero. The effect is more significant when FS is included. This points to the fact that the $\Lambda$CDM model is a robust fit to these data.
  
  \item Comparing datasets with and without FS, we see that the effect of adding the FS dataset shifts $H_0$ towards the Planck value. The shift is about $1.4\sigma$ without SH0ES, and $1.3\sigma$ with SH0ES.
  
  \item A residual tension still remains when the SH0ES constraint on $H_0$ is included.
  Without FS, the fits show a $1.5\sigma$ tension in the $H_0$ posterior, while with FS, the tension increases to $2.6\sigma$.
  
  \item As we show in Tables~\ref{tab:PlanckRnR} and~\ref{tab:PlanckFSRnR}, the $\chi^2$ of the fit including the EDE model does not improve relevantly with respect to $\Lambda$CDM for any experiment but for SH0ES.
  Even in the case of SH0ES, when we add the FS, the improvement is much decreased.
\end{itemize}


\section*{Acknowledgements}
\noindent On May 21st 2020, we presented our results to the Simons' Modern Inflationary collaboration meeting.

\noindent We thank Vivian Poulin for comments on the draft and Martin Sloth  for discussions. We thank Francis-Yan Cyr-Racine for sharing and help in running his RnR code, Evan McDonough, Colin Hill and Michael Toomey for sharing the CLASS EDE code, and Vivian Poulin for sharing and help in running the AxiCLASS code.
LS is partially supported by the Simons Foundation Origins of the Universe program (Modern Inflationary Cosmology collaboration) and by NSF award 1720397. 
Part of the analysis was performed on the Sherlock cluster at the Stanford University, for which we thank the support team, and part on the HPC (High Performance Computing) facility of the University of Parma, whose support team we thank.

\appendix

\section{Shift of the Sound Horizon}\label{app:shift}

\begin{figure}[ht]
  \includegraphics[width=0.54\linewidth]{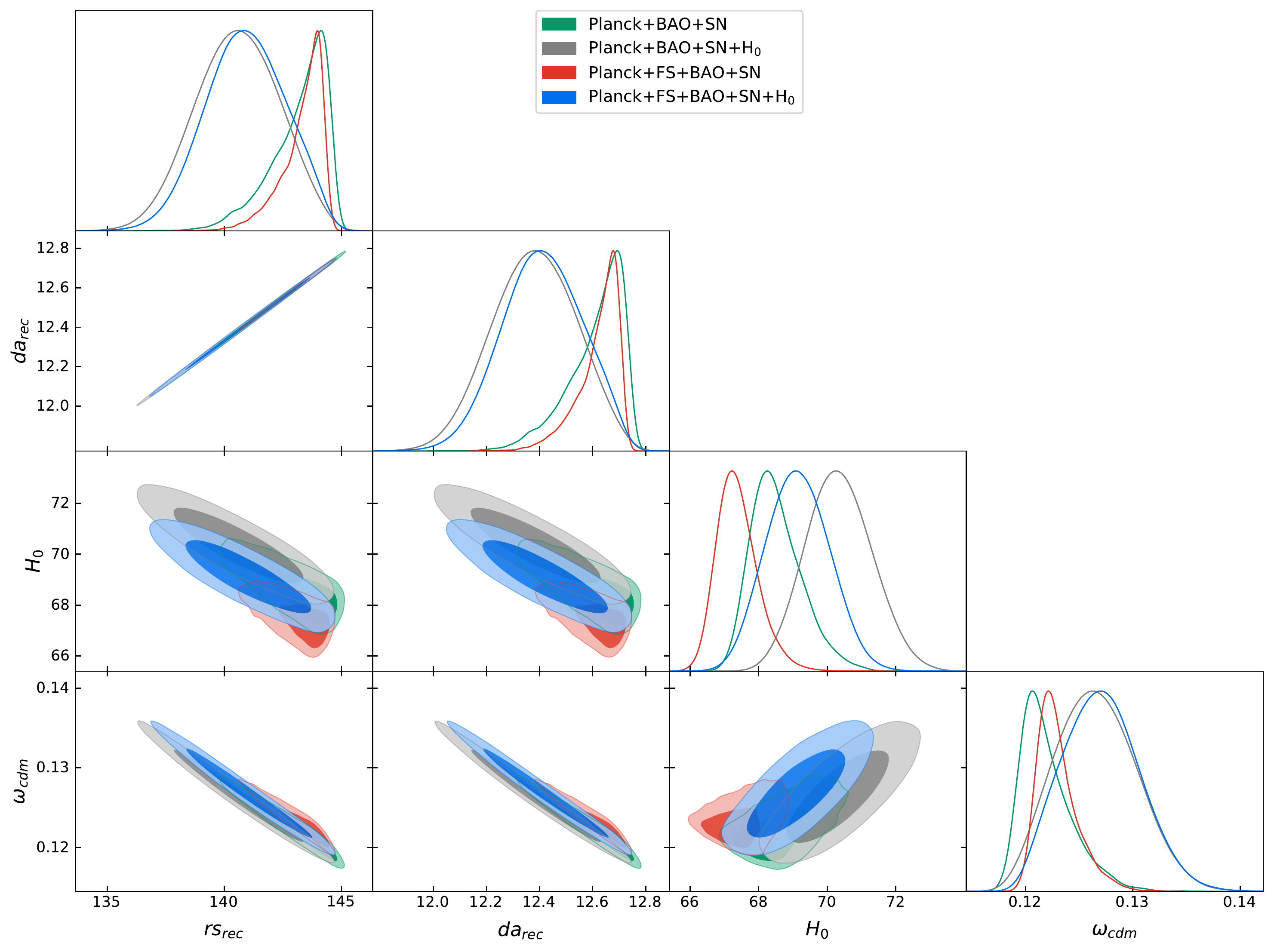}
  \includegraphics[width=0.54\linewidth]{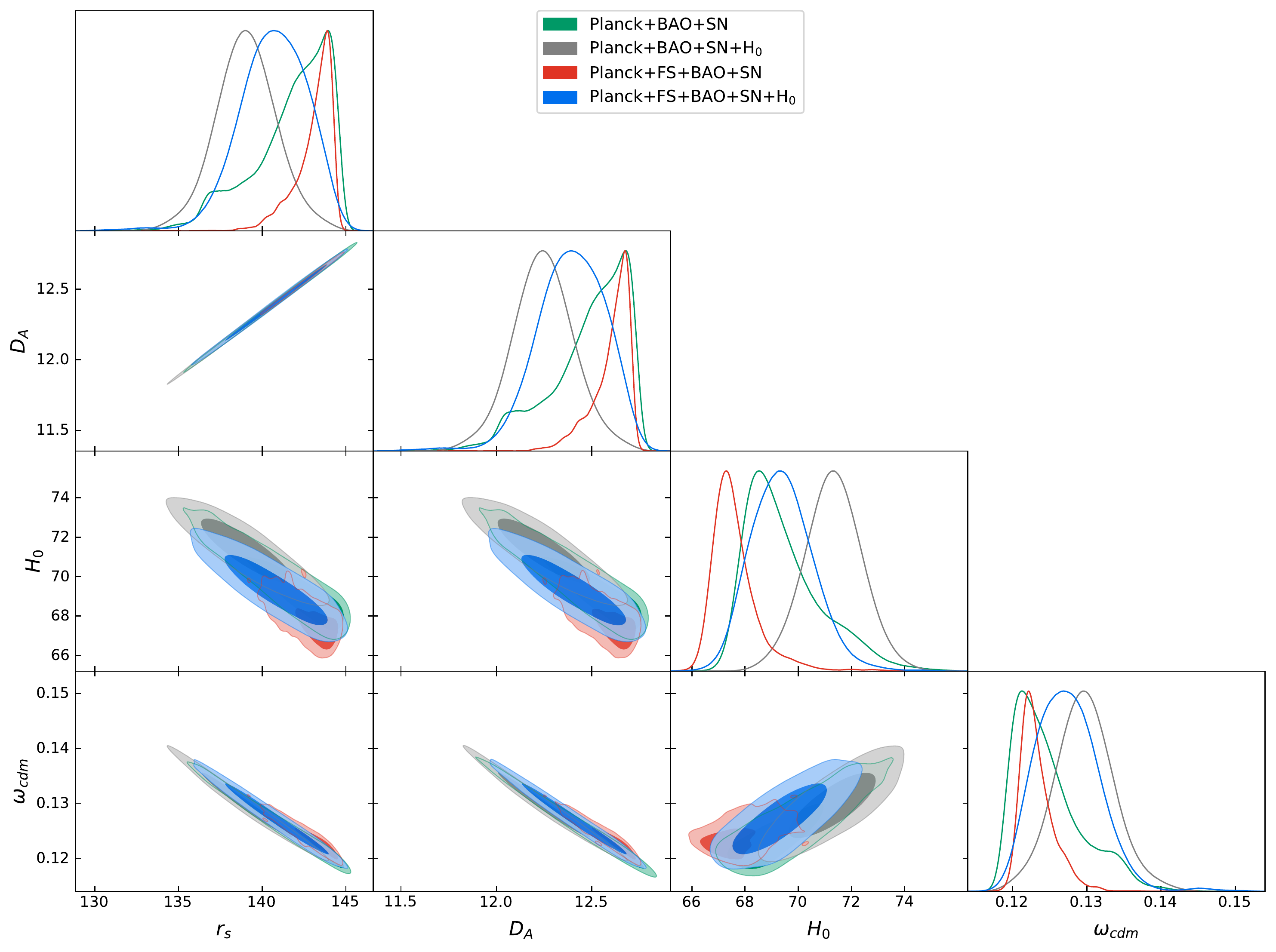}
  \caption{Plots of the sound horizon $r_s$, angular {diameter} distance to recombination $D_A$, $H_0$ and $\omega_{cdm}$ for the four different dataset combinations analyzed.
  \emph{Left:} RnR model. \emph{Right:} EDE model.}
  \label{fig:rs_plot}
\end{figure}

As an illustration of the discussion in sec.~\ref{sec:review}, in Fig.~\ref{fig:rs_plot} we show the fits to the sound horizon and the angular diameter distance at recombination, together with $H_0$ and the cosmological constant parameters, for the four different combinations of datasets in the RnR and EDE models.

The degeneracy line in the $r_s - D_A$ plane, corresponding to fixed $\theta_s$, is manifest.
We can also see the inverse relation between $r_s$ and $H_0$, and $D_A$ and $H_0$ which are exploited, {or at least attempted to exploit}, by the early dark energy models. {As anticipated by the physical considerations in sec.~\ref{sec:review}, the combination of Planck and LSS data (as well as, in these plots, of Supernovae data) allows for a rather precise determination of $r_s$ close to the one of the Planck $\Lambda$CDM model: $147.19\pm 0.29$ Mpc. This implies that $H_0$ is not allowed to move much from the $\Lambda$CDM fit, even within the freedom in $r_s$ theoretically allowed for by the EDE and RnR models.}

\section{Comparison with $\Lambda$CDM and FS Information}
\label{sec:lcdm_comp}
\begin{figure}[!ht]
  \includegraphics[scale=0.42]{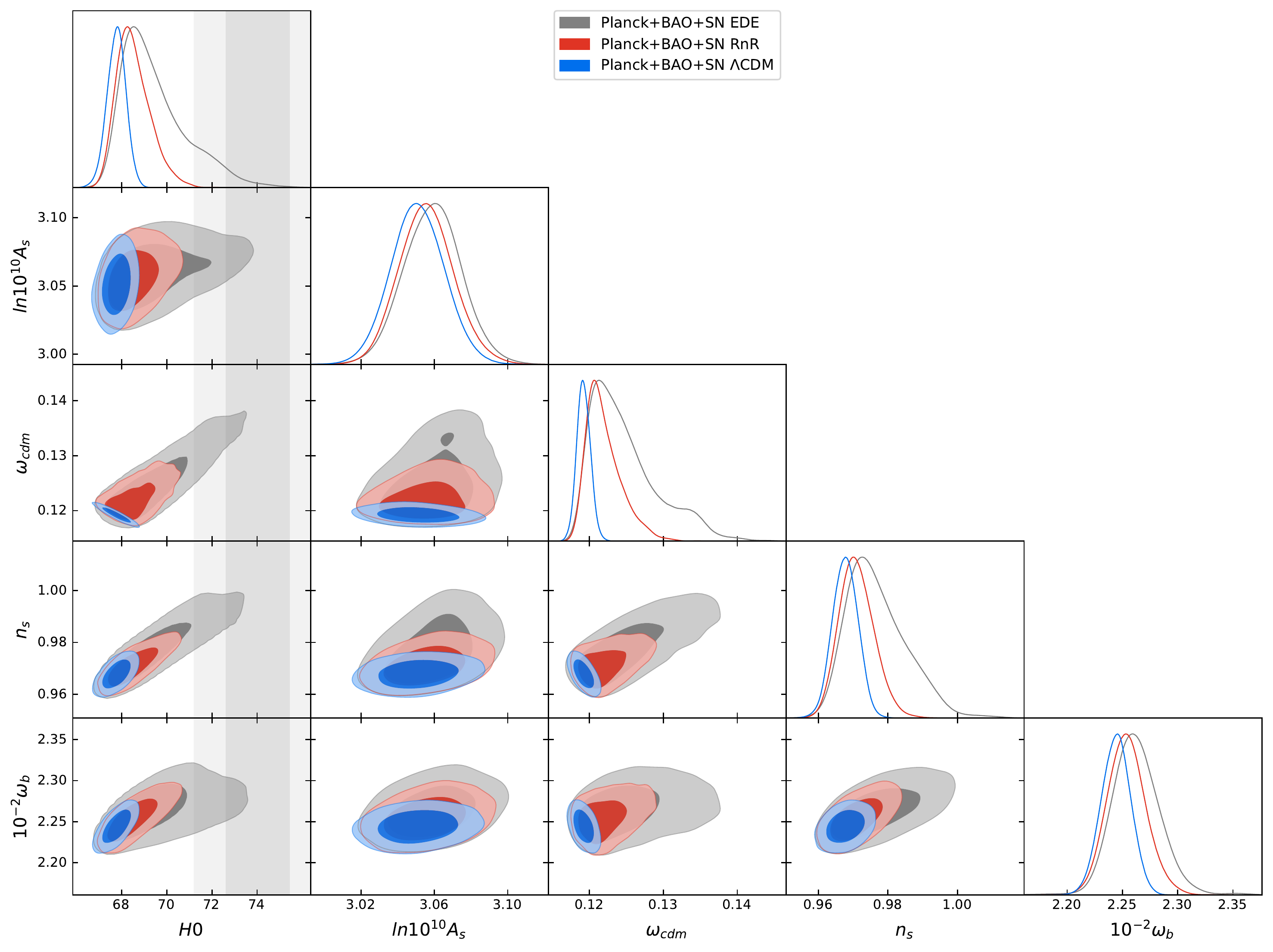}
  \caption{Plots of the cosmological parameters in the three models for the Planck+BAO+SN dataset.}
  \label{fig:planck_noH0}
\end{figure}

\begin{figure}[!ht]
  \includegraphics[scale=0.42]{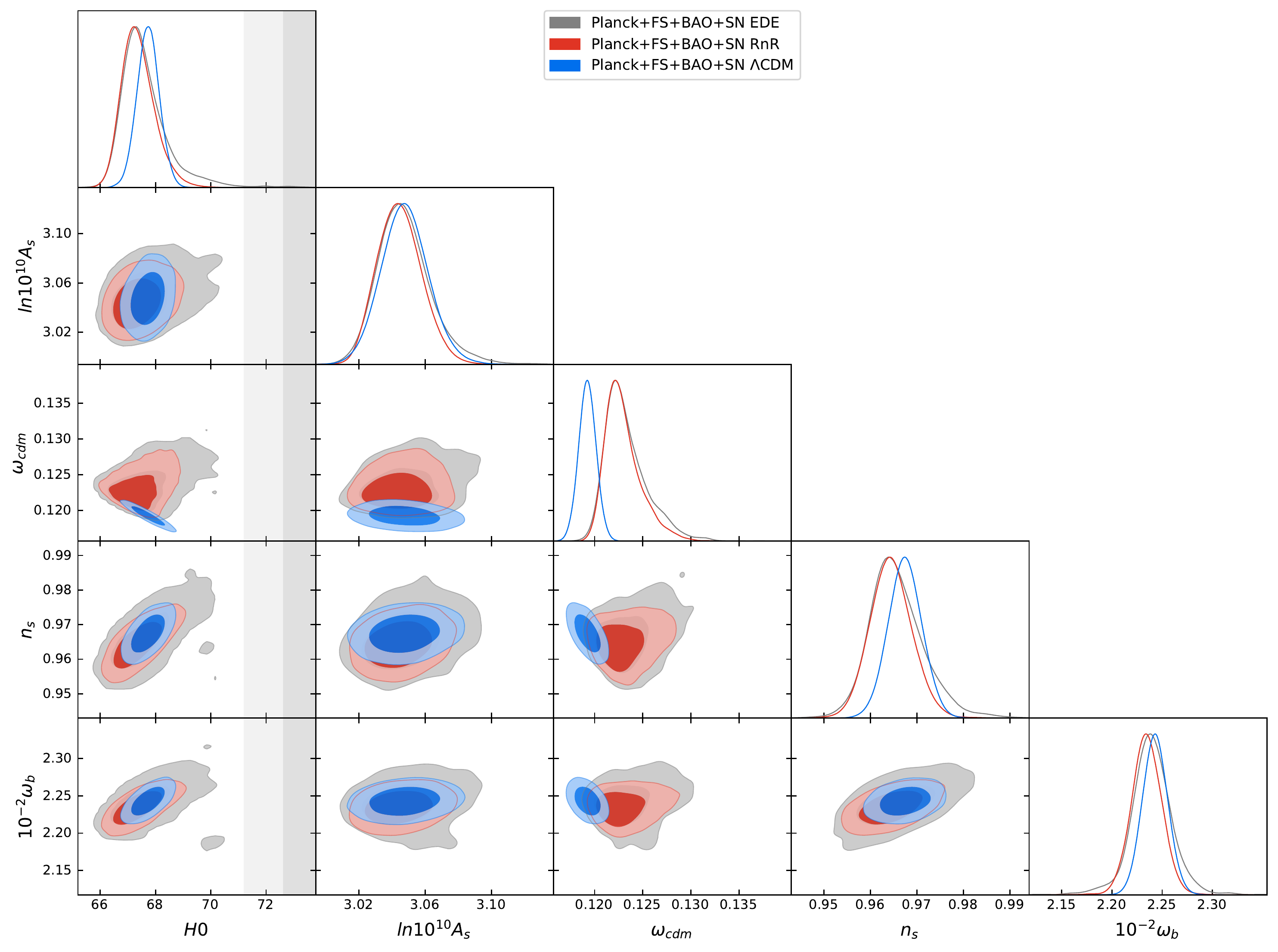}
  \caption{Plots of the cosmological parameters in the three models for the Planck+FS+BAO+SN dataset.}
  \label{fig:planckfs_noH0}
\end{figure}

To check that our runs are consistent, here we show the posteriors for their common cosmological parameters, compared with the $\Lambda$CDM model run on the same datasets.
As it is apparent from Fig.~\ref{fig:planck_noH0} and \ref{fig:planckfs_noH0}, the concordance is quite good, except for a little shift in the $\omega_{cdm}$ parameter due to effect of early dark energy on the shape of the matter power spectrum. Explaining this allows us to highlight the physical effects of the EDE and RnR models (or in general of models that make the sound horizon decrease) on the dark matter power spectrum.

The presence of the early dark energy makes the time during which the gravitational potential decays in radiation domination shorter, leading to larger peak amplitude, which can be compensated by a larger $\omega_m$, as from~eq.~(\ref{eq:peaks}) (see also~\cite{Poulin:2018cxd,Smith:2019ihp}). Notice that this shift in $\omega_m$ gives rise to a larger broad band part of the dark matter power spectrum (see also~\cite{Hill:2020osr}). In fact, if $k_{\rm eq}$ is the wavenumber that is equal to the horizon at equality, $k_{\rm eq}$ grows with $\omega_m$, and the dark matter power spectrum decays as roughly $(k_{\rm eq}/k)^2$ for $k\gtrsim k_{\rm eq}$ (see for example~\cite{DAmico:2019fhj}), and so it grows with larger~$\omega_m$. This growth in the broad band is compensated by quite a large shift in the tilt of the power spectrum, which is still very compatible with Planck, but, without FS, was bluer than Planck, but still compatible with it. Also, as from eq.~(\ref{eq:thetacmb}) and~(\ref{eq:BAOtotal}), a  larger $\omega_m$ increases the acoustic angles in the CMB and in LSS, which is compensated by a decrease in $H_0$, which is also visible in the plot.


\section{$\chi^2$-tables}\label{app:chisquare}

In Tables~\ref{tab:PlanckRnR} and \ref{tab:PlanckFSRnR}, we present the best-fit $\chi^2$ for each experiment in the runs including and without including the FS information.

\begin{table}[!h]
  \begin{tabular}{l|lll|lll|}
  \cline{2-7}
                                                    & $\Lambda$CDM & EDE   & RnR   & $\Lambda$CDM & EDE   & RnR   \\ \hline
  \multicolumn{1}{|l|}{Planck high-$\ell$ TT+TE+EE} & 585.5        & 583.6 & 584.0 & 587.0        & 584.3 & 590.4 \\ \hline
  \multicolumn{1}{|l|}{Planck low-$\ell$ EE}            & 396.6        & 396.9 & 397.2 & 397.1        & 397.2 & 397.5 \\ \hline
  \multicolumn{1}{|l|}{Planck low-$\ell$ TT}            & 22.86        & 21.02 & 22.84 & 22.72        & 21.07 & 21.53 \\ \hline
  \multicolumn{1}{|l|}{Planck lensing}             & 8.952        & 9.249 & 8.871 & 9.127        & 9.407 & 9.136 \\ \hline
  \multicolumn{1}{|l|}{BAO BOSS DR12}             & 4.293        & 3.629 & 4.252 & 3.525        & 3.427 & 3.622 \\ \hline
  \multicolumn{1}{|l|}{Pantheon}                    & 1027         & 1027  & 1027  & 1027         & 1027  & 1027  \\ \hline
  \multicolumn{1}{|l|}{BAO small-z 2014}           & 1.257        & 1.580 & 1.274 & 1.633        & 2.001 & 1.650 \\ \hline
  \multicolumn{1}{|l|}{eBOSS DR14 Ly-$\alpha$}  & 4.9          & 4.67  & 4.88  & 4.67         & 4.43  & 4.58  \\ \hline
  \multicolumn{1}{|l|}{SH0ES}                       &              &       &       & 16.97        & 2.002 & 5.753 \\ \hline
  \multicolumn{1}{|l|}{Total}                       & 2052         & 2048  & 2051  & 2070         & 2051  & 2061  \\ \hline
  \end{tabular}
\caption{Best-fit $\chi^2$ for each experiment in the runs without the FS, without SH0ES in the left half and with it in the right half.}
\label{tab:PlanckRnR}
\end{table}

\begin{table}[!h]
  \begin{tabular}{l|lll|lll|}
  \cline{2-7}
                                                    & $\Lambda$CDM & EDE   & RnR   & $\Lambda$CDM & EDE   & RnR   \\ \hline
  \multicolumn{1}{|l|}{Planck high-$\ell$ TT+TE+EE} & 585.0        & 582.9 & 585.5 & 588.3        & 584.7 & 587.3 \\ \hline
  \multicolumn{1}{|l|}{Planck low-$\ell$ EE}            & 396.6        & 396.9 & 395.8 & 396.9        & 396.1 & 395.9 \\ \hline
  \multicolumn{1}{|l|}{Planck low-$\ell$ TT}            & 23.10        & 22.23 & 23.99 & 22.60        & 21.51 & 22.74 \\ \hline
  \multicolumn{1}{|l|}{Planck lensing}             & 8.855        & 9.01 & 8.818 & 9.715        & 9.500 & 9.050 \\ \hline
  \multicolumn{1}{|l|}{BOSS DR12 FS + BAO, high-z NGC}      & 57.67        & 58.42 & 59.13 & 57.37        & 59.43 & 59.29  \\ \hline
  \multicolumn{1}{|l|}{BOSS DR12 FS + BAO, high-z SGC}      & 68.87        & 69.50 & 68.34 & 69.67        & 69.69 & 68.17 \\ \hline
  \multicolumn{1}{|l|}{BOSS DR12 FS + BAO, low-z NGC}       & 62.49        & 62.44 & 62.12 & 63.27        & 62.45 & 62.14  \\ \hline
  \multicolumn{1}{|l|}{Pantheon}                    & 1027         & 1027  & 1029  & 1027         & 1027  & 1028  \\ \hline
  \multicolumn{1}{|l|}{BAO small-z 2014}           & 1.162        & 1.392 & 0.741 & 1.90         & 1.156 & 0.845  \\ \hline
  \multicolumn{1}{|l|}{eBOSS DR14 Ly-$\alpha$}  & 4.97         & 4.79  & 5.5   & 4.54         & 4.92  & 5.2  \\ \hline
  \multicolumn{1}{|l|}{SH0ES}                       &              &       &       & 15.90        & 6.289 & 13.41 \\ \hline
  \multicolumn{1}{|l|}{Total}                       & 2236         & 2235  & 2239  & 2257         & 2243  & 2252   \\ \hline
  \end{tabular}
\caption{Best-fit $\chi^2$ for each experiment in the runs including the FS, without SH0ES in the left half and with it in the right half.}
\label{tab:PlanckFSRnR}
\end{table}

\section{Checks on Volume Sampling Effects}
\label{app:sampling}
Both the analysis for the EDE  and the RnR models may be affected at some level by a degeneracy in parameter space that occurs when the energy injection due to scalar field goes to zero.
In this regime, all the other model parameters  do not play any physical role, so that the model gives the same predictions as $\Lambda$CDM.
In this sense, there may be an artificially large volume in parameter space associated to vanishing $f_{\rm EDE}$ or $f_{\rm RnR}$ that might enhance the statistical weight of the $\Lambda$CDM models when using some sampling algorithms such as Metropolis-Hastings, as pointed out, for example, in~\cite{Smith:2019ihp,Lin:2019qug,Niedermann:2019olb}.

\begin{figure}[ht]
  \includegraphics[width=0.54\linewidth]{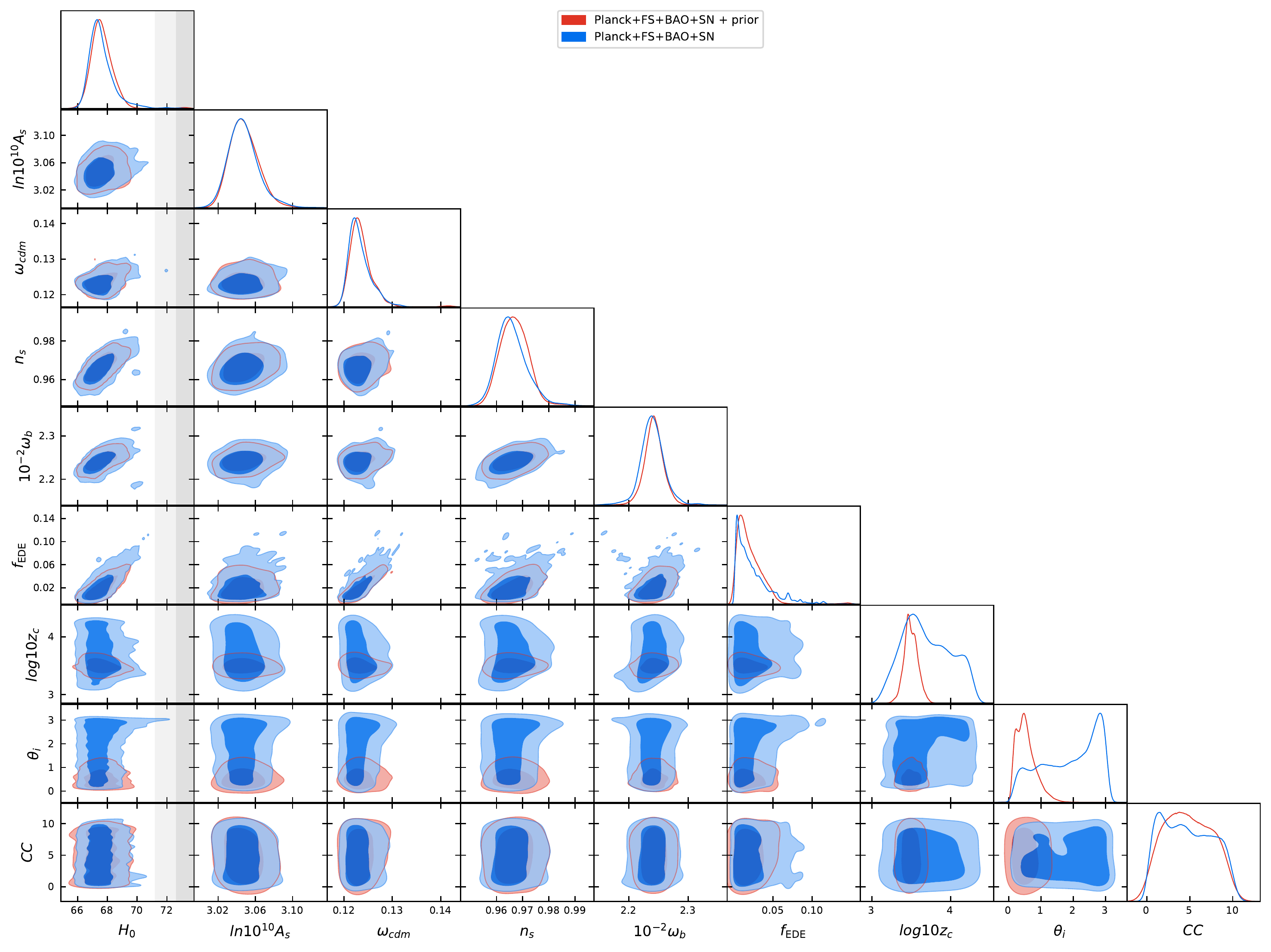}
  \includegraphics[width=0.54\linewidth]{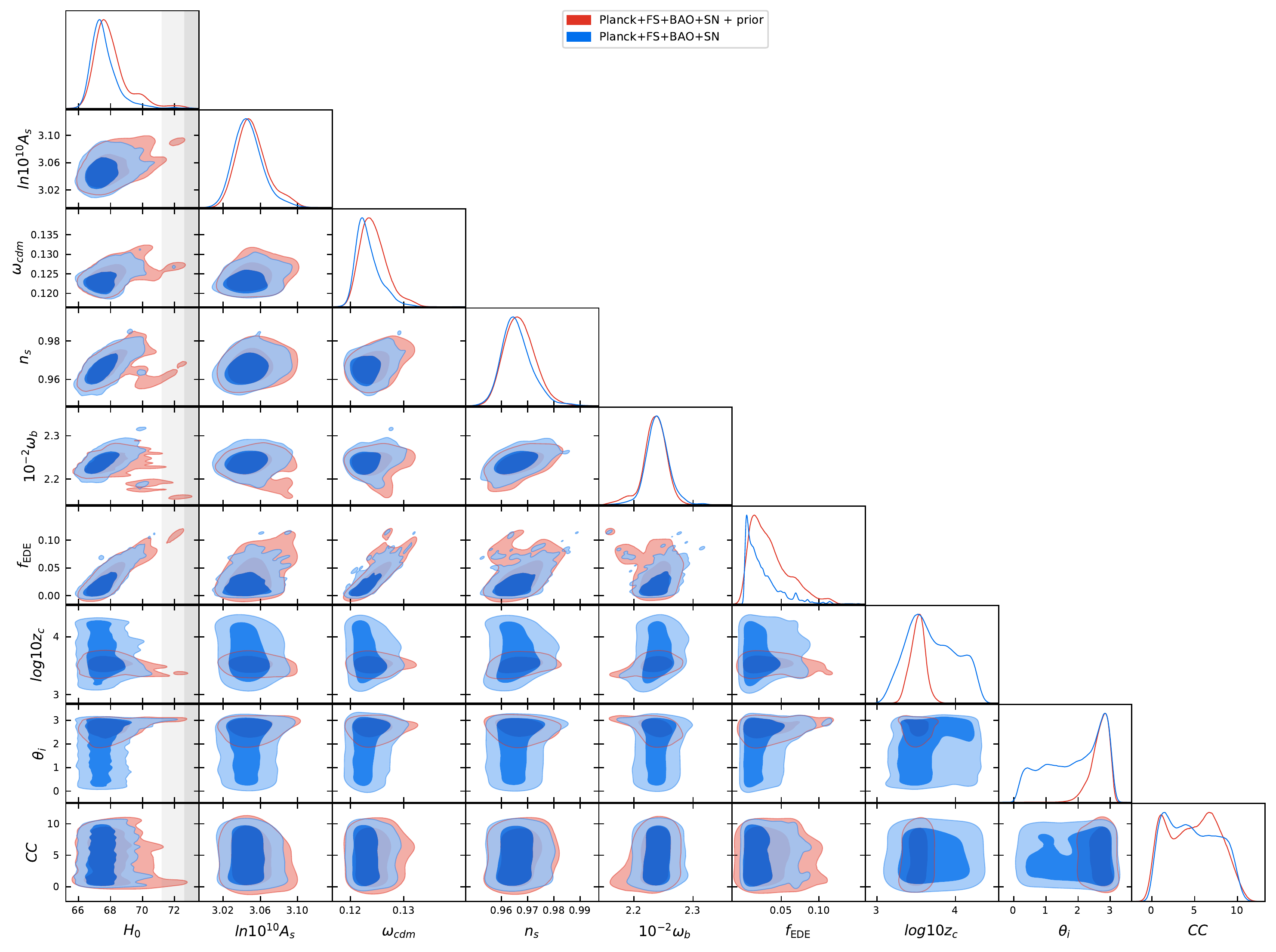}
  \caption{\emph{Left:}  Posterior distribution for EDE for the combination Planck+FS+BAO+SN, before ({\it blue}) and after ({\it  red}) imposing a prior on $z_c$ to be close to recombination and $\Theta_i$ to be small. \emph{Right:} Same as left, but imposing a prior on $z_c$ to be close to recombination and $\Theta_i$ to be large.}
  \label{fig:ede_IS}
\end{figure}

In order to address to what extent this affects our findings, we perform the following importance sampling of our MCMC, similarly to what done in~\cite{Niedermann:2019olb}.
We start with the EDE model, because in  this case the degeneracy in parameter space is larger: for $f_{\rm EDE}=0$, both $z_c$ and $\Theta_i$ do not play any role.
We also just focus on the combination of Planck+FS+BAO+SN data, which is the most constraining set that is affected by this potential issue~\footnote{When SH0ES data are included, the region $f_{\rm EDE}\approx 0$ is hardly sampled.}.
We first impose a Gaussian prior on the logarithm of the injection redshift, $\log_{10} z_c$, close to the best fit: average 3.5 and standard deviation 0.1.
In this way, the degeneracy associated to $z_c$ is removed.
Then, on the left of Fig.~\ref{fig:ede_IS}, we present our posteriors after imposing an additional prior on $\Theta_i$ to be small~\footnote{We impose a half-Gaussian prior with standard deviation of $0.5$ and mean $0.1$, which is the lower bound on $\Theta_i$ of the original chain.}, compared to our results in the main text.
We see that the distribution of $f_{\rm EDE}$ has barely changed, becoming  slightly  wider away  from zero.
The posterior for the cosmological parameters have  hardly  changed.
This shows that {\it  most} of the sampling points of the likelihood associated with $f_{\rm EDE}\approx 0$ has actually $z_c$ close to recombination, and a small initial field value, implying a physically small energy injection.
Vice-versa, on the right of Fig.~\ref{fig:ede_IS}, we now impose the same prior on $z_c$ and an additional prior on $\Theta_i$ to be large~\footnote{Here impose a half-Gaussian prior with standard deviation of $0.5$ and mean $3.1$, which is the upper bound on $\Theta_i$ of the original chain.}.
This prior allow us to focus on the parameter region where  the energy is made to vanish by sending the parameter $f$ of the axion generating the EDE to zero.
This is the only other way to send the energy injection to zero in EDE beyond sending $\Theta_i$ to zero.
In this case, the posterior of $f_{\rm EDE}$ is enhanced away from zero, as expected, but it is still strongly peaked towards zero, suggesting that even in this case the data prefer no energy injection at all.
The posteriors for the cosmological parameters are hardly changed as well.
We therefore conclude that, for EDE and within our analysis, volume sampling issues play a negligible role.

\begin{figure}[ht]
  \centering
  \includegraphics[scale=0.4]{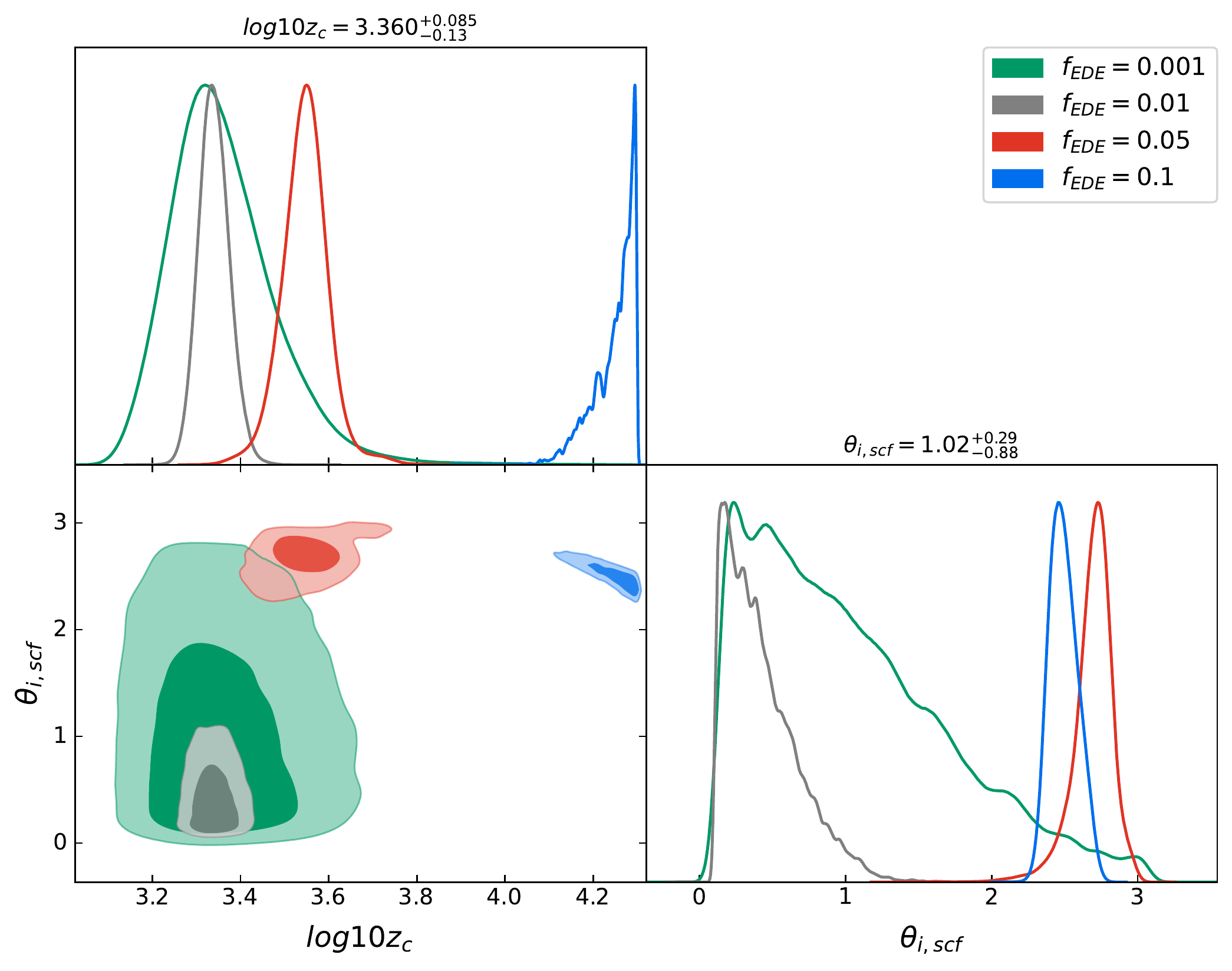}
  \caption{Posterior distributions for the combination Planck+FS+BAO+SN in the $\theta_i-z_c$ subspace, for 4 different values of $f_{EDE}$, fixing all other parameters to their best fit values.}
  \label{fig:fede_zc_th}
\end{figure}

We do another check of the volume sampling effects, in the following way.
We consider the combination Planck+FS+BAO+SN, and fix all the parameters to their best fit values, except $f_{EDE}$, $\Theta_i$ and $z_c$.
Then, we run 4 different MCMC in the subspace $\Theta_i-z_c$ fixing $f_{EDE}$ to $0.001, 0.01, 0.05, 0.1$ (as a reference, the best fit for the full chain is $f_{EDE}=0.0522$).
The results are shown in fig.~\ref{fig:fede_zc_th}.
We see that small values of $f_{EDE}$ result in a posterior peaked at $z_c$ close to recombination, and $\Theta_i$ close to zero, which corresponds to a physically small energy injection. In particular, we do not observe a flat posterior that would imply a large volume sampling effect, questioning our analysis.
Larger values of $f_{EDE}$ give a posterior in $\Theta_i$ peaked towards larger initial angles, which means there is a larger energy density in the scalar field.
The posterior of $z_c$ moves to very high values in the case $f_{EDE} = 0.1$, since for such large EDE fractions we have to lower the energy density in the EDE close to recombination, otherwise one could not have a good fit to the CMB.
These results are therefore consistent with the importance sampling performed above.

\begin{figure}[ht]
  \centering
  \includegraphics[width=0.84\linewidth]{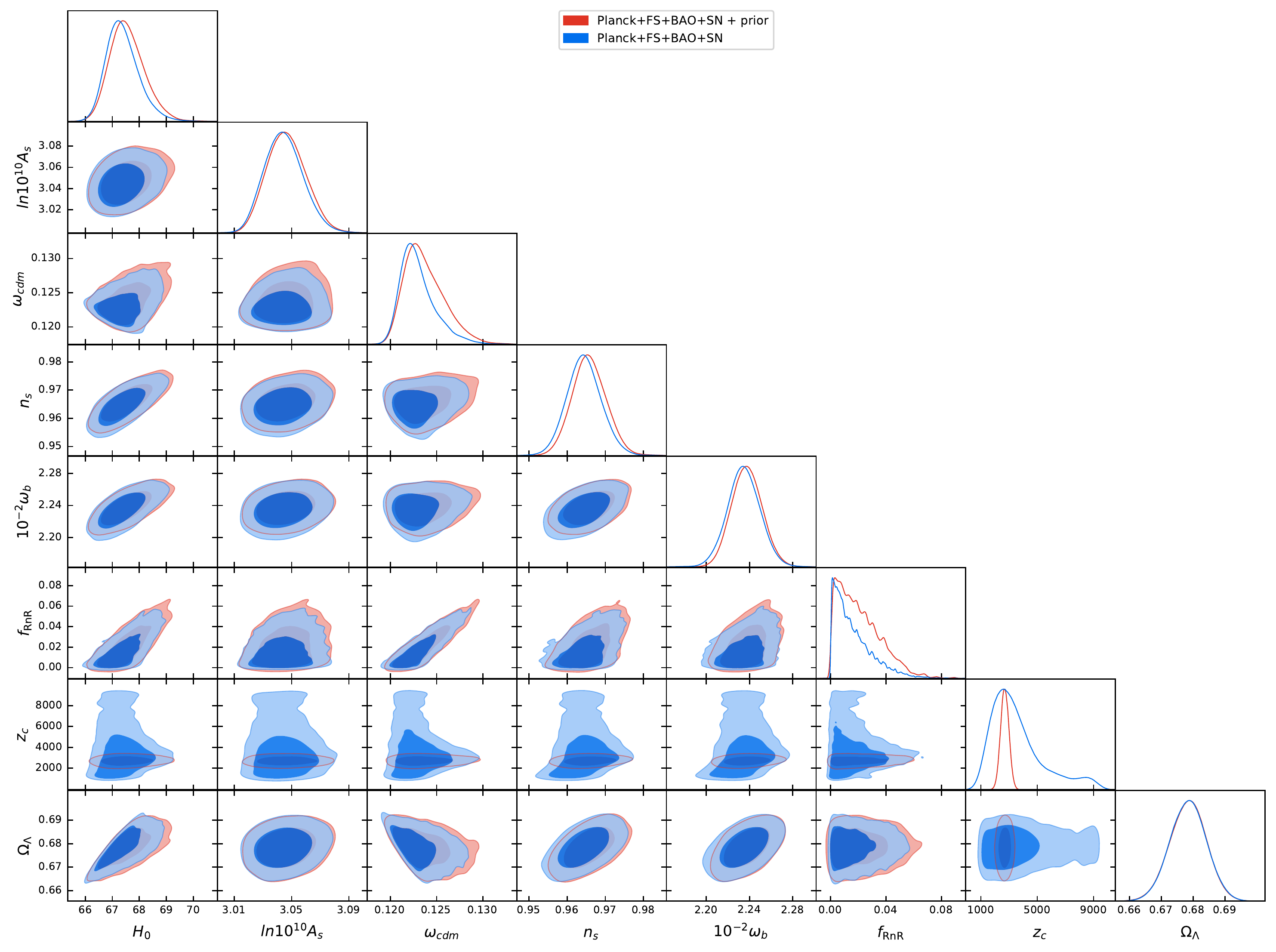}
  \caption{Posterior distribution for RnR for the combination Planck+FS+BAO+SN, before ({\it blue}) and after ({\it  red}) imposing a prior on $z_c$ to be close to recombination.}
  \label{fig:rnr_IS}
\end{figure}

We now pass to similarly analyze the RnR model. Here the smaller number of parameters of the model makes the  treatment more straightforward: to check for the  volume sampling effects, we just need to impose  the  prior on $z_c$ to be close to recombination. We choose a Gaussian prior with mean $2700$ and standard deviation 300, close to the peak of the distribution.
The resulting posteriors are shown in Fig.~\ref{fig:rnr_IS}.
The conclusions are the same: while the $f_{\rm RnR}$ distribution is slightly enhanced away from zero, it is still very strongly peaked towards zero, implying that the data actually prefer no energy injection at all.
The distribution of the other parameter are hardly changed as well.
Therefore, also in the RnR model and within our analysis, volume sampling issues play a negligible role.

\section{Effect of different power spectrum amplitude}
\label{app:Alss}
One question that may be raised is whether it is consistent to combine the BOSS power spectrum and the CMB dataset in a single analysis, since there may be tensions between the datasets, which in turn could drive the fit of $H_0$ (see for example~\cite{Smith:2020rxx}).
In particular, the results of~\cite{DAmico:2019fhj} show that the FS-only analysis prefers a value for the amplitude of the primordial power spectrum, $\ln(10^{10} A_s)$, about 1.8 $\sigma$ lower than the best-fit Planck value.

\begin{figure}[ht]
  \centering
  \includegraphics[width=0.84\linewidth]{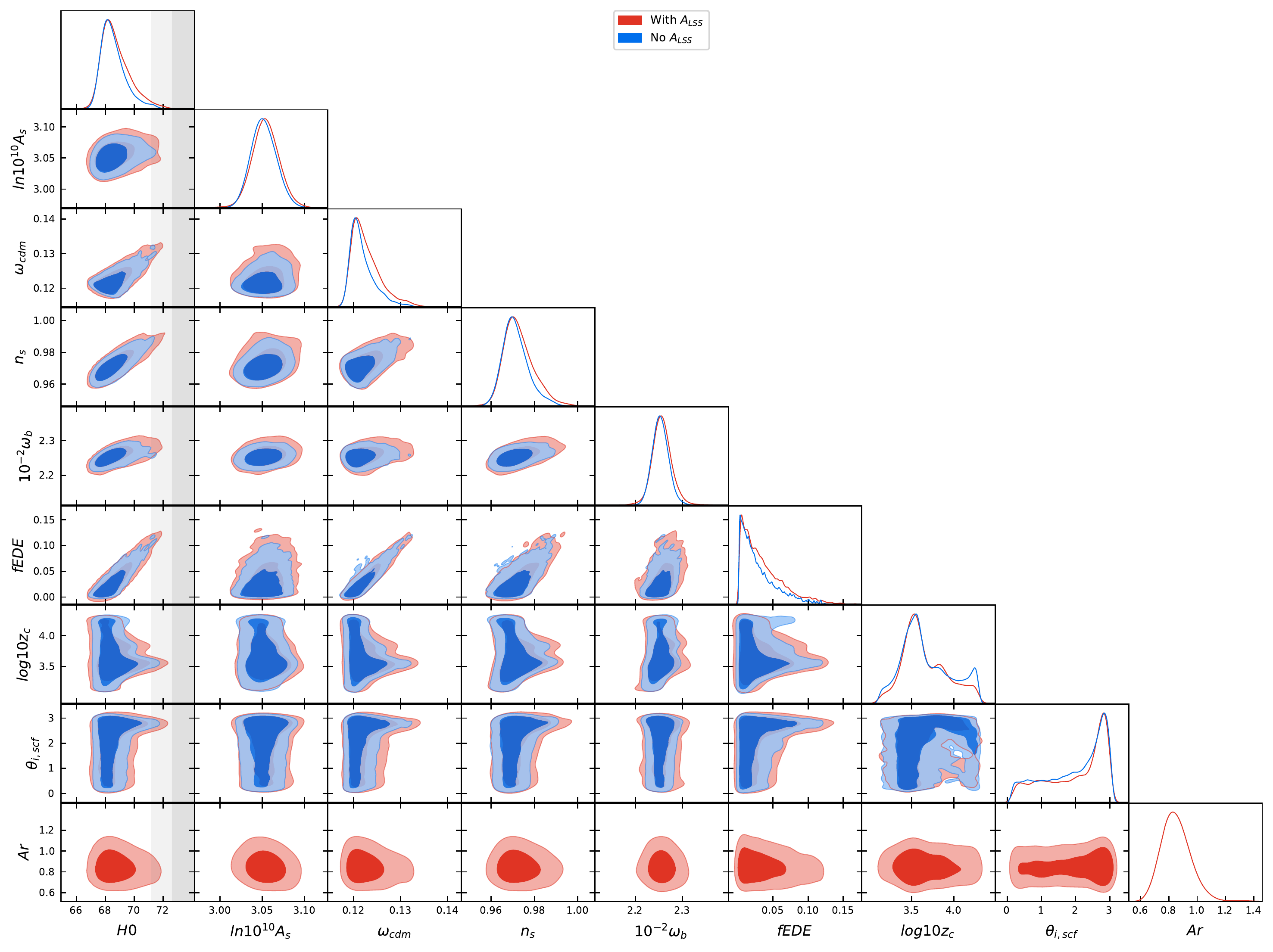}
  \caption{Posterior distribution for the combination Planck+FS+BAO+SN, including ({\it red}) or not ({\it  blue}) the parameter $A_r$ parametrizing the ratio of linear power spectrum used in the FS likelihood to the one used in the Planck likelihood.}
  \label{fig:Alss}
\end{figure}

We check for this possible issue by fitting the combination Planck+FS+BAO+SN adding a parameter, $A_r$, which is the ratio of the amplitude of the linear power spectrum used in the FS likelihood to the one used in the Planck likelihood.
Our results are shown in fig.~\ref{fig:Alss}.
While we determine $A_r = 0.852^{+0.086}_{-0.12}$ at $68\%$ CL (less than 2$\sigma$ away from 1), the posterior distribution for the parameters of the EDE model, and in particular $H_0$, are not appreciably shifted.
Explicitly, in the run with $A_r$, we find $H_0 = 68.75^{+0.60}_{-1.2}$, while in the run without it we find $H_0 = 68.57^{+0.48}_{-1.0}$, a shift of less than a third of a sigma.
We conclude that the analysis combining the FS dataset with the Planck one is consistent, and the difference in the best-fit values for the amplitude of the primordial power spectrum has no impact on the EDE fit.
We can see extremely minor shifts in the parameters.

\bibliographystyle{JHEP}
\bibliography{references}

\end{document}